\documentclass{article}

\usepackage{microtype}
\usepackage{graphicx}
\usepackage{enumitem}
\usepackage{subcaption}
\usepackage{float}
\usepackage{hyperref}
\usepackage{booktabs}

\usepackage{amsmath}
\usepackage{amssymb}
\usepackage{mathtools}
\usepackage{amsthm}
\usepackage[table,xcdraw]{xcolor}

\usepackage[accepted]{mlsys2025}

\newcommand{\niparagraph}[1]{\smallskip\noindent\textbf{#1}}

\newcommand{\customlabelname}{position}
\newcommand{\customautoref}[1]{\hyperref[#1]{\customlabelname}}

\newcommand{\systemname}{ExpertFlow}

\begin{document}
\twocolumn[

\mlsystitle{\systemname: Adaptive Expert Scheduling and Memory Coordination for Efficient MoE Inference}


\mlsyssetsymbol{equal}{$^{\dagger}$}
\mlsyssetsymbol{correspond}{$^{*}$}

\begin{mlsysauthorlist}
\mlsysauthor{Zixu Shen}{equal,to}
\mlsysauthor{Kexin Chu}{equal,to}
\mlsysauthor{Yifan Zhang}{to}
\mlsysauthor{Dawei Xiang}{to}
\mlsysauthor{Runxin Wu}{to}
\mlsysauthor{Wei Zhang}{correspond,to}
\end{mlsysauthorlist}

\mlsysaffiliation{to}{School of Computing, University of Connecticut, CT, USA}
\mlsyscorrespondingauthor{Wei Zhang}{wei.13.zhang@uconn.edu}


\renewcommand{\mlsysEqualContribution}{}

\mlsyskeywords{Mixture-of-Experts (MoE), Cache-Aware Routing, Scheduling, LLM inference Optimization}
\vskip 0.3in
\begin{abstract}
The expansion of large language models is increasingly limited by the constrained memory capacity of modern GPUs. To mitigate this, Mixture-of-Experts (MoE) architectures activate only a small portion of parameters during inference, significantly lowering both memory demand and computational overhead. However, conventional MoE inference approaches, which select active experts independently at each layer, often introduce considerable latency because of frequent parameter transfers between host and GPU memory. In addition, current cross-layer prediction strategies, which are typically based on fixed steps, lack adaptability across different hardware platforms and workloads, thereby reducing their robustness and effectiveness.

To address these challenges, we present \systemname, a runtime system for MoE inference that combines adaptive expert prefetching and cache-aware routing. \systemname\ continuously adjusts its prediction horizon for expert activation by leveraging runtime statistics such as transfer bandwidth, parameter dimensionality, and model feedback signals. Furthermore, it incorporates a hybrid cross-layer prediction scheme that fuses pregating information with intermediate computational states to anticipate future expert needs. By adaptively refining prefetching decisions and aligning them with actual usage behavior, \systemname\ effectively decreases cache misses and removes latency caused by expert swap-ins. Our evaluation demonstrates that \systemname\ reduces model stall time to less than 0.1\% of the baseline, highlighting its capability to optimize MoE inference under stringent memory constraints.
\end{abstract}
]



\printAffiliationsAndNotice{\mlsysEqualContribution} 

\section{Introduction}
\label{sec:intro}
The growing complexity of large language models has placed increasing emphasis on architectural designs that offer both scalability and efficiency. Among these, Mixture-of-Experts (MoE) has become a dominant architecture for scaling large language models, owing to its sparse activation strategy~\cite{shazeer2017outrageously}. By activating only a small subset of expert subnetworks for each input, MoE models can scale to billions of parameters without incurring a proportional computational cost. This principle has enabled breakthroughs in various high-profile systems, including Switch Transformer~\cite{fedus2022switch}, GShard~\cite{lepikhin2020gshard}, GPT-3~\cite{brown2020language}, and GPT-4.

\begin{figure}
    \centering 
    {\includegraphics[width=0.99\linewidth]{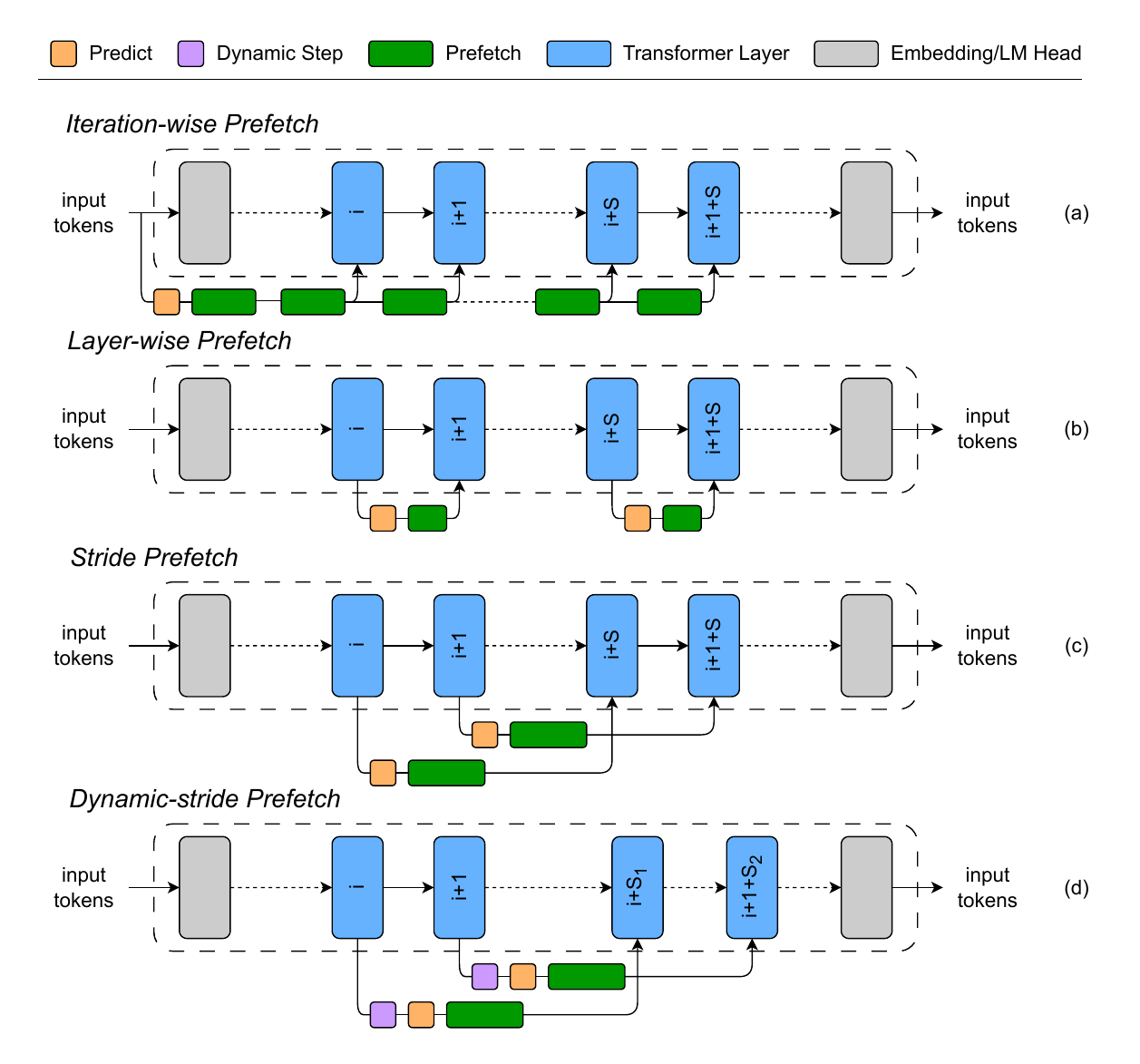}}
    \vspace{-0.5em}
    \caption{Candidate prefetch manners}
\label{fig:overview}
\vspace{-1.0em}
\end{figure}

As deployment of MoE models continues to expand, the need for responsive and resource-aware expert routing mechanisms has become increasingly urgent. To this end, existing models commonly adopt one of three strategies, as illustrated in~\autoref{fig:overview}(a)--(c). In some systems, the expert assignment is performed statically, each layer is bound to a fixed set of experts prior to execution, and that assignment does not change at runtime. Others adopt a reactive approach, where expert selection for a given layer depends on the activations from the preceding layer. A third option employs periodic scheduling: the current layer predicts expert configurations for a fixed number of layers ahead, reducing routing frequency at the cost of flexibility.

Although each of these strategies brings computational benefits, they also suffer from fundamental limitations. Static expert assignment leads to linear growth in memory and compute overhead as model depth increases~\cite{shazeer2017outrageously, cao2025moe}. Per-layer reactive prediction neglects global coordination, often resulting in noisy or inconsistent expert usage. Meanwhile, fixed-interval forecasting improves efficiency by amortizing routing cost, but is inherently inflexible and ill-suited for handling diverse input dynamics.

To address these issues, we propose \systemname, a runtime system for MoE inference that combines adaptive expert prefetching and cache-aware routing. As shown in~\autoref{fig:overview}(d), it adaptively determines how far in advance expert predictions should be made within the model. Our approach introduces a token-aware scheduling policy, which effectively fuses fine-grained input features, such as token-level statistics, with internal routing scores to guide the appropriate depth of expert forecasting. Unlike prior work that relies on a fixed lookahead window applied uniformly across all inputs, our system instead adjusts the routing interval dynamically on a per-input basis, thereby enabling both greater accuracy and improved system-wide efficiency.

Our contributions are threefold: 
\begin{itemize}[leftmargin=1.5em, itemsep=1pt, topsep=2pt]
    \item \textbf{Adaptive Cross-layer Prefetch:} We propose an algorithm that adaptively selects the depth at which expert predictions are performed, achieving a trade-off between inference cost and model fidelity. This design aligns with the broader trend of adaptive computation~\cite{graves2016adaptive, wang2018skipnet}. 
    \item \textbf{Token-Routing Integration:} We design a mechanism that merges conventional MoE routing signals with token-level metadata, enhancing both the precision and reliability of expert activation. 
    \item \textbf{Hierarchical Expert Coordination:} Drawing inspiration from hierarchical mixtures of experts~\cite{716791}, we introduce a multi-level expert scheduling framework that governs expert reuse and cooperation across various depths of the model.
\end{itemize}

Together, these contributions deliver a more intelligent and flexible expert routing solution for MoE models. Our framework not only improves routing quality and consistency but also significantly reduces computational overhead, facilitating scalable deployment of MoE-based inference in diverse and resource-constrained environments.

\section{Background and Motivation}
\subsection{MoE Models}
MoE architectures scale large language models by activating only a small subset of experts for each input~\cite{shazeer2017outrageously, fedus2021switch, yang2024xmoe}. This sparse execution achieves high parameter capacity without proportional computational cost~\cite{zhou2022mixture}. Yet, most MoE systems rely on a fixed expert prediction interval (known as step size $S$), which determines how many tokens are processed before rerouting. While easy to implement, this static interval overlooks both hardware and workload variability.

A constant $S$ cannot accommodate changes in GPU bandwidth, memory pressure, or token-level routing divergence. Step sizes tuned for one platform often yield suboptimal utilization on another, and static tuning fails when batch statistics fluctuate across inputs. Moreover,  $S$ embodies an inherent latency–accuracy tension: small values incur frequent expert switches and overhead, whereas large values degrade routing precision and inflate memory usage. Consequently, existing MoE designs remain inefficient under dynamic runtime conditions, motivating adaptive step-size mechanisms that adjust routing granularity based on real-time workload and platform feedback.

\begin{figure}[h]
    \centering 
    {\includegraphics[width=0.99\linewidth]{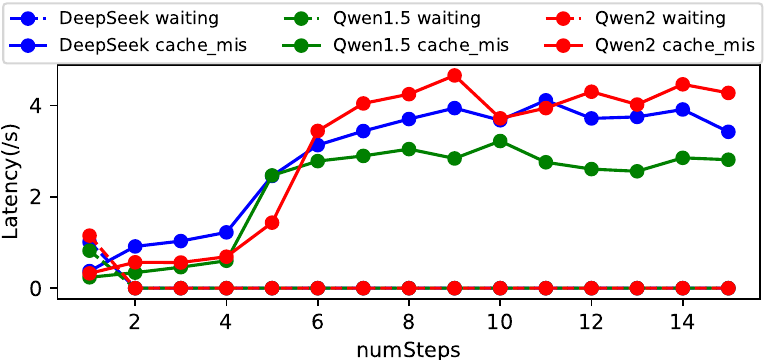}}
    \vspace{-0.5em}
    \caption{Latency Comparison Across the Step Size ($S$)}
\label{fig:M1}
\vspace{-1.0em}
\end{figure}

\subsection{Motivation}

Dynamic step-size scheduling fundamentally shapes the balance between computation and communication in MoE inference. As shown in \autoref{fig:M1}, both large and small step values degrade throughput for distinct reasons. Larger steps widen the prediction horizon for prefetching, lowering prediction accuracy and increasing cache misses that waste GPU cycles. Smaller steps, in contrast, lead to synchronization stalls as computation frequently waits for expert loading, especially in deep MoE stacks where latency accumulates layer-by-layer. The following analysis identifies the primary factors that influence step-size selection.

\paragraph{\textbf{Hardware Bandwidth Sensitivity:}} 
As shown in \autoref{fig:M1}, on A6000 GPUs, setting \texttt{numSteps}=2 effectively removes most of the waiting latency. This behavior is governed by hardware bandwidth. During inference, MoE computation depends on expert parameters that must be loaded before execution begins. The total step size can be estimated based on model size, available bandwidth, and per-layer computation time. Generally, GPUs with higher PCIe bandwidth can overlap multiple layers of computation with expert transfers, while slower interconnects, such as those on RTX 4090 require larger \texttt{numSteps} to achieve comparable overlap (~\autoref{tab:gpu_bandwidth}).

\noindent
\colorbox{gray!15}{\parbox{0.97\linewidth}{
\textbf{Observation I:} The optimal step size depends on hardware characteristics. Different GPU bandwidths necessitate distinct scheduling strategies to balance computation and communication overlap.
}}

\begin{table}[htp]
\centering
\vspace{-1.0em}
\caption{GPU and Transfer Bandwidths}
\label{tab:gpu_bandwidth}
\begin{tabular}{c c}
\hline\hline
\textbf{GPU} & \textbf{Bandwidth (GB/s)} \\ \hline
NVIDIA H20         & 128                               \\
ASCEND 910B        & 128
\\
NVIDIA A100        & 64                               \\ 
NVIDIA A6000       & 64                               \\ 
NVIDIA RTX 4090    & 32                               
\\
Intel Arc B580     & 16                               \\ 
AMD Radeon RX 6500 XT   &8                             \\ \hline\hline
\end{tabular}
\end{table}

\begin{figure}
    \centering 
    {\includegraphics[width=0.99\linewidth]{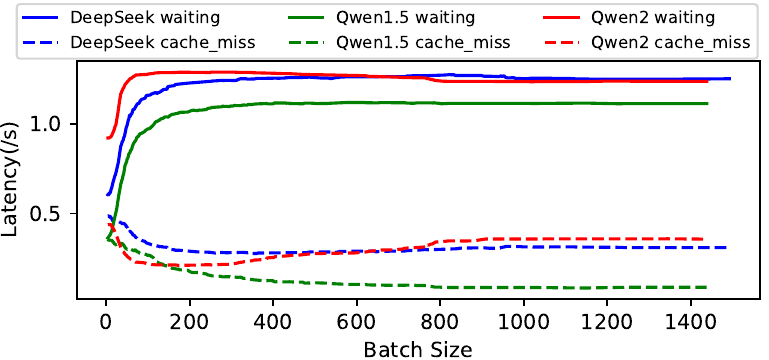}}
    \vspace{-0.5em}
    \caption{Latency comparison across batch sizes.}
    \label{fig:M3}
    \vspace{-0.5em}
\end{figure}

\paragraph{\textbf{Workload Scale and Contention:}} 
Even under fixed hardware, workload parameters significantly affect latency. As shown in \autoref{fig:M3}, increasing batch size initially improves throughput by amortizing expert-loading overhead. Past a certain point, however, larger batches intensify contention for shared GPU memory and I/O bandwidth. This raises token-level diversity and enlarges the total candidate expert set \(N_{\text{total}}\), which in turn stresses the swap-in path. When preloaded experts align with available bandwidth, cache misses fall and latency drops. When they do not align, redundant expert loading amplifies end-to-end delay. This results in a non-monotonic trend: after batch size enters a saturation region, average waiting latency increases by up to 2.5\(\times\), even though cache-miss latency decreases by 12–40\%, following
\[
L_{\text{cache}} \propto R_{\text{miss}} = 1 - \frac{N_{\text{selected}}}{N_{\text{total}}},
\]
where \(N_{\text{selected}}\) and \(N_{\text{total}}\) denote correctly predicted and total experts. These results indicate that a single static step size cannot capture the interaction between workload scale, GPU memory pressure, and cache behavior.

\noindent
\colorbox{gray!15}{\parbox{0.97\linewidth}{
\textbf{Observation II:} Workload scale influences latency through both resource contention and expert diversity. 
}}

\begin{figure}
    \centering 
    {\includegraphics[width=0.99\linewidth]{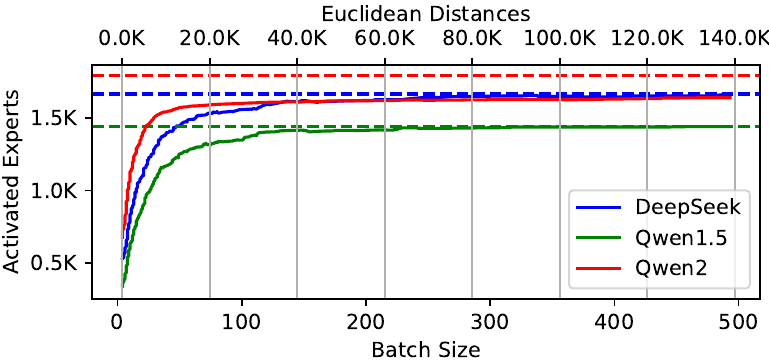}}
    \vspace{-0.5em}
    \caption{Variation of activated experts with batch size and Euclidean distance}
\label{fig:M5}
\vspace{-0.5em}
\end{figure}

\paragraph{\textbf{Intra-Batch Diversity:}} 
Batch size alone does not fully explain expert activation behavior. Tokens within the same request may trigger very different experts depending on their semantics.  To quantify this diversity, we define the cumulative Euclidean distance among token embeddings as:
$$
\text{Dist}(\mathbf{t}) = \sum_{1 \le i < j \le k} \|\mathbf{v}_{t_i} - \mathbf{v}_{t_j}\|_2,
$$
where \(\mathbf{v}_{t_i}\) is the embedding of token \(t_i\). This metric measures how widely the token representations are spread in embedding space, and it correlates more strongly with expert demand than batch size alone. Empirically, requests with the same batch size can vary in end-to-end latency by up to 300\% solely due to differences in semantic diversity. As shown in \autoref{fig:M5}, \(\mathrm{Dist}(\mathbf{t})\) grows smoothly with workload complexity and provides a stable signal for anticipating expert reuse, swap pressure, and cache residency.

\noindent
\colorbox{gray!15}{\parbox{0.97\linewidth}{
\textbf{Observation III:} Semantic diversity within a batch is a reliable predictor of expert demand. Cumulative embedding distance offers a stable runtime signal for adaptive step-size selection and cache-aware scheduling.
}}

\noindent
Collectively, these observations motivate a dynamic, bandwidth- and input-aware scheduling mechanism that jointly considers hardware bandwidth, workload scale, and token diversity. Such adaptive scheduling minimizes cache misses and synchronization stalls, achieving stable and efficient MoE inference across diverse hardware and workload configurations.

\subsection{Challenges in Implementing the Method}

Implementing dynamic step size control in MoE architectures presents several non-trivial challenges that must be addressed to ensure both execution efficiency and system reliability.

\subsubsection{\textbf{Dynamic Step Size Selection.}}\ 
At the start of each iteration, the cumulative Euclidean distances among tokens in the current batch are computed to estimate expected expert activations per layer. Deriving a suitable step size \(S\) involves multiple interacting factors, including expert module size, device bandwidth, and per-layer latency. For example, the expert loading cost can be approximated as:
$$
\text{swap\_in\_latency} = \frac{\text{predicted\_num\_experts} \times \text{SIZE}}{\text{bandwidth}},
$$
which reflects the overhead from expert swaps. The step size should balance this latency against layer compute time to ensure timely prefetching. This becomes more complex under varying workload patterns and hardware conditions, requiring a method that remains robust across different runtime scenarios.

\subsubsection{\textbf{Expert Prediction Reliability.}}\ 
After determining \(S\), the next challenge is ensuring accurate expert preselection to maximize cache efficiency. Standard pre-gating—passing the hidden state directly to the next-layer router—often lacks predictive reliability. A more robust mechanism is needed, ideally incorporating richer context across layers and potentially leveraging learned models to better capture expert selection patterns. Improving prediction fidelity is critical for reducing cache misses and aligning preselected experts with actual usage.

\subsubsection{\textbf{Real-time Step Size Adjustment.}}\ 
Even with a well-chosen initial \(S\), runtime conditions may necessitate adjustments. If expert wait times accumulate due to underestimation, the system should be able to expand \(S\) on the fly. This requires a feedback loop that monitors indicators like cache miss rate and wait time, and adjusts \(S\) responsively. The control logic must be stable and responsive, avoiding oscillations or overcompensation that could disrupt inference performance.

\section{System Design}
\label{sec:design}
\subsection{Overview}

The key insight of \systemname\ is to dynamically determine a hardware-aware adaptive step size \(S\) for cross-layer expert activation in Mixture-of-Experts (MoE) models. Instead of statically activating experts layer-by-layer, \systemname\ predicts the required experts several layers in advance and continuously updates \(S\) based on runtime conditions. This adaptive process prevents the cumulative latency that typically arises from sequential expert swapping, aligning expert activation with GPU memory availability and interconnect bandwidth.

\begin{figure}
    \centering 
    {\includegraphics[width=0.99\linewidth]{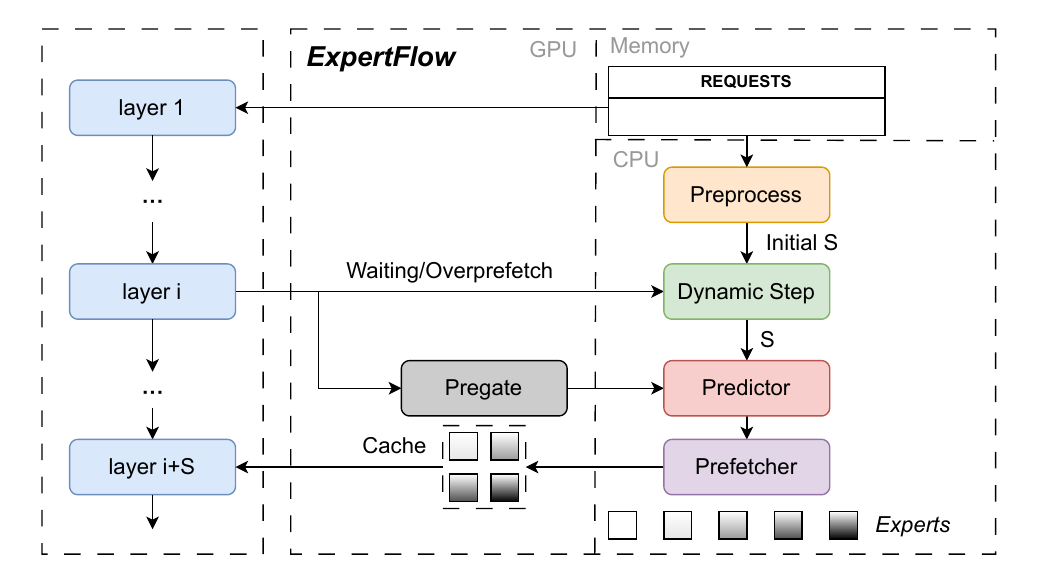}}
    \vspace{-0.5em}
    \caption{The System Architecture of \systemname}
\label{fig:architecture}
\vspace{-1.0em}
\end{figure}

As shown in~\autoref{fig:architecture}, to achieve accurate and efficient prediction, \systemname\ combines a pre-gate strategy with a predictive mechanism that utilizes both token identifiers and recent expert activation states derived from hidden representations. This dual-source design improves the accuracy of expert prefetching, minimizes cache misses, and sustains high GPU utilization. The initial value of \(S\) is determined by expert size, transfer bandwidth, and token dependency distance, and is later refined through real-time monitoring of waiting latency. By integrating pre-gate outputs with token semantics and activation history, \systemname\ ensures consistent accuracy under dynamic workloads. Together, these techniques enable \systemname\ to deliver high-throughput MoE inference with reduced computational and communication overhead.

\subsection{Adaptive Expert Pre-gating}
\subsubsection{\textbf{Dynamic Expert Activation.}}
\begin{figure}[t]
    \centering 
    {\includegraphics[width=0.99\linewidth]{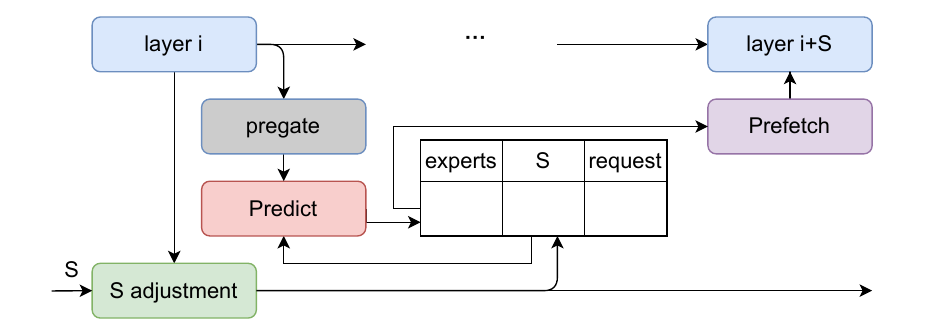}}
    \vspace{-0.5em}
    \caption{Mechanism for Determining Dynamic Step Size (\(S\)).}
\label{fig:codesign}
\vspace{-1.0em}
\end{figure}

\systemname\ improves the efficiency of MoE models by dynamically determining the number of layers to skip (\(S\)) based on communication cost, expert characteristics, and device capabilities. The mechanism is illustrated in~\autoref{fig:codesign}.

The process begins by estimating the number of experts likely to be activated per layer. This is derived from pre-gate outputs, which provide a probability distribution over experts. Experts are selected in descending order until their cumulative probability surpasses a predefined threshold, yielding an expected activation count. Combined with expert size and device communication bandwidth, this informs the calculation of \(S\), the number of layers to skip.

The step size is formally computed as:
\[
S = \frac{N_e \cdot E_s}{C_s \cdot T_l},
\]
where \(N_e\) is the number of experts to activate, \(E_s\) is expert size, \(C_s\) is communication bandwidth, and \(T_l\) is per-layer compute time.

This formulation adapts \(S\) to reflect the trade-off between communication overhead and computation. During inference, \systemname\ updates \(S\) in real time using feedback from prediction accuracy. When predictions degrade, \(S\) increases to reduce reliance on uncertain forecasts; when predictions remain stable, \(S\) is lowered to maximize prefetching efficiency.

\subsubsection{\textbf{Cross-Layer Prediction and Delta Adjustment.}}
Cross-layer prediction is a core component of \systemname, enabling adaptive expert activation based on historical patterns and real-time feedback. A pre-trained model, built from past pre-gate outputs and actual expert activations, learns to estimate their deviation($\Delta$), which is added to the pre-gate result to refine predictions for future layers.

This model relies on three inputs: \textit{(1). pre-gate output of the current layer, as a baseline estimate}; \textit{(2). token IDs of the current batch, capturing input-specific semantics}; \textit{(3). expert activation history from previous layers, reflecting inter-layer dependencies}. By integrating these signals, the model corrects naive pre-gate predictions, enhancing expert selection accuracy and reducing unnecessary activations.

To maintain runtime efficiency, \systemname\ first checks for cached predictions. If a cached prediction for that (token sequence, layer index, and step size) exists, \systemname\ reuses it. Otherwise, the system falls back to direct top-k selection from the router logits. This dual-mode design ensures efficient yet reliable routing.

\systemname\ also employs a lightweight feedback loop to adapt step size \(S\). During execution, if a predicted expert is unavailable (causing wait latency), a stall counter increases. When it exceeds a threshold, the counter resets and \(S\) is incremented by 1. Conversely, if experts are already loaded when needed, an overfetch counter increases. Once this reaches its threshold, it resets and decrements \(S\) by 1. This loop allows \(S\) to self-adjust based on prediction effectiveness and expert readiness, balancing latency and accuracy.

\subsubsection{\textbf{File Collection and Data Parsing.}}
We first collect the activation metadata of the current MoE model, which includes the token ID list, corresponding layer indices, the list of predicted experts, the list of actually activated experts during execution, and the adaptive step size \(S\). After validation and parsing, each line is converted into a structured sample:
\[
\text{Sample}_i = \left\{
\begin{array}{l}
\text{token\_ids}: \mathbf{t}_i,\\
\text{layer\_idx}: l_i,\\
\text{predicted\_experts}: \mathbf{p}_i,\\
\text{actual\_experts}: \mathbf{a}_i,\\
S: s_i
\end{array}
\right\},
\]
and the dataset is denoted by \(\mathcal{D} = \{\text{Sample}_i\}_{i=1}^n\).

This structure provides a clean representation of expert activation behavior at each layer and step configuration, which serves as the foundation for downstream grouping and learning.

\subsubsection{\textbf{Request Grouping and Feature Construction.}}
Samples are grouped by the tuple \((\mathbf{t}, S)\), i.e., token ID sequences and step size, under the assumption that each group represents a unique input request under a specific step configuration:
\[
\mathcal{G} = \{(\mathbf{t}, S) \mapsto \{\text{Sample}_j \mid \text{token\_ids} = \mathbf{t},\ S = s_j\}\}.
\]

For each group, we construct training examples by iterating over layers in ascending order. Token sequences are first embedded via a fixed random embedding table \(\mathbf{E} \in \mathbb{R}^{V \times d}\), and aggregated by mean pooling:
\[
\mathbf{e} = \frac{1}{k} \sum_{i=1}^k \mathbf{E}_{t_i},
\]
where \(V\) is the vocabulary size and \(d\) is the embedding dimension.

Each training feature vector is constructed as:
\[
\mathbf{x} = [\mathbf{e}, S, l, \text{prev\_act}],
\]
where \(l\) is the current layer index, and \(\text{prev\_act} \in \{0, 1\}^{L \cdot M}\) is a binary vector encoding the activation status of experts in all previous layers, assuming \(L\) total layers and \(M\) experts per layer.

The corresponding label is a one-hot vector indicating the actual experts used in the current layer:
$$\mathbf{y} \in \{0, 1\}^M.$$ 
By constructing one training example per layer per request, we capture temporal activation dependencies across layers and steps.

\subsubsection{\textbf{Training and Evaluation.}}
The final dataset consists of feature-label pairs:
\[
X \in \mathbb{R}^{N \times F}, \quad Y \in \mathbb{R}^{N \times M},
\]
where \(N\) is the number of samples, \(F = d + 2 + L \cdot M\) is the feature dimensionality, and \(M\) is the number of experts per layer.

We use the RandomForestRegressor model from the \texttt{scikit-learn} library~\cite{scikit-learn}. We selected it because it runs on CPU, introduces negligible GPU interference, and can be evaluated asynchronously with respect to GPU kernels. We also found that shallow MLP predictors offered comparable accuracy but consumed GPU compute during inference, which directly contends with model execution.

The model is trained to minimize Mean Squared Error:
\[
\text{MSE} = \frac{1}{N} \sum_{i=1}^N \|\mathbf{y}_i - \hat{\mathbf{y}}_i\|^2,
\]
where \(\mathbf{y}_i\) is the true expert activation vector and \(\hat{\mathbf{y}}_i\) is the model's prediction.

To evaluate prediction quality, we compute the accuracy as the proportion of correctly predicted expert bits, averaged across all samples. Additionally, to capture how performance varies under different execution configurations, we report per-step-size accuracy for each value of \(S\).

This end-to-end pipeline, from parsing expert activation logs to learning with a random forest model, enables the system to anticipate expert usage patterns based on token semantics, prior activations, and layer context. The resulting predictor supports accurate expert prefetch planning, reducing cache misses and latency during MoE inference.

\subsection{Expert Memory Management}

Effective memory management is critical for sustaining high throughput during MoE inference, particularly when GPU memory is limited. \systemname\ introduces an advanced two-tiered caching architecture, complemented by dynamic reassignment and coordinated integration with the prefetch module, to optimize expert residency and minimize costly reloads.

\subsubsection{\textbf{Two-Level LRU Cache}}
Rather than relying on a single eviction queue, \systemname\ maintains two distinct LRU structures:
\begin{itemize}[leftmargin=1.5em, itemsep=1pt, topsep=2pt]
\item \texttt{LRU\_high}: Stores experts with demonstrated high reuse potential. These typically include experts accessed within the last  layers or those predicted for imminent activation by the expert prediction module.
\item \texttt{LRU\_low}: Contains experts that have not been recently accessed or are less likely to be reused in the near future.
\end{itemize}
Evictions are performed preferentially from \texttt{LRU\_low}, which offloads completed or low-priority experts to DRAM and preserves high-reuse experts in GPU memory. In contrast to prior single-level LRU strategies, this two-tiered approach preserves frequently accessed experts in GPU memory while prioritizing less critical ones for eviction. This design reduces memory pressure, allowing the system to scale effectively to larger expert pools and operate efficiently under constrained GPU memory conditions.

A key motivation for this design stems from the sequential nature of layer-wise MoE inference. Under standard eviction policies, experts from the initial \(S\) layers—though often reused later during decoding—are prematurely evicted, resulting in frequent cache misses and increased latency.

As the step size \(S\) dynamically evolves throughout execution, expert-to-tier assignments are continuously updated. This ensures that the caching policy remains aligned with the current inference path and maintains near-optimal memory utilization over time.

\subsubsection{\textbf{Prefetch and Cache Coordination}}
Memory management and dynamic activation operate in tandem: 
\textit{(1). Prior to computing the skip distance \textit{S}, \systemname\ queries current cache occupancy to estimate effective bandwidth and anticipated eviction cost}. 
\textit{(2). Post-prefetch, observed transfer times update the bandwidth estimate $C_s$, informing subsequent skip-distance calculations}. 
\textit{(3). Feedback counters for prefetch adaptation incorporate cache hit/miss statistics, ensuring that memory decisions reinforce activation strategies rather than conflict with them}.
This feedback-driven coordination aligns expert loading with cache availability, preventing redundant load/evict cycles and reducing overall data movement.

\subsection{Cache-aware Routing}
Cache misses disrupt MoE execution by introducing asynchronous stalls that delay computation. Unlike fixed-point waiting latency, cache misses occur unpredictably, but this very asynchrony creates an opportunity for latency hiding through task overlap.

To exploit this, \systemname\ overlaps cache miss handling with ongoing computation. When a miss triggers an expert swap-in, the transfer proceeds in parallel with active inference tasks. A cache-aware routing policy gives scheduling priority to tokens whose required experts are already resident in memory, while deferring those that would cause additional I/O.

In contrast, conventional MoE systems perform swap-in sequentially and often block due to contention with swap-out operations. \systemname\ addresses this by assigning cache miss resolution the highest priority in the memory queue, ensuring timely fault handling and improving overall throughput by reducing exposed latency.
\section{Evaluation}
\label{sec:evaluation}


In this section, we perform an extensive evaluation of \systemname\ and demonstrate its performance improvement in terms of reduced waiting latency and improved MoE expert prediction accuracy. We also test the impact of request length, and different MoE vectors on the performance of \systemname. In our evaluation, we aim to answer the following questions:

\begin{enumerate}[leftmargin=1.5em, itemsep=1pt, topsep=2pt]
    \item Compared with the baseline, what effects does \systemname\ exert on waiting latency and cache miss latency? (\autoref{overall_latency_declines})
    \item Compared with the baseline, how does \systemname\ affect expert activation prediction accuracy?  (\autoref{pre-gate-based Predictor})
    \item Compared with the baseline, is the proposed swap-out policy effective? (\autoref{Memory_Management_Policy})
    \item Is the cache-aware routing used in \systemname\ effective? (\autoref{Cache-aware Routing})
\end{enumerate}

\subsection{Experiment Setup}

\niparagraph{Testbeds. }
All experiments were conducted independently on three GPU platforms: NVIDIA A6000~\cite{nvidiaA6000datasheet} (64 GB/s bandwidth), H20 (128 GB/s), and Ascend 910B (128 GB/s). \systemname\ was implemented in Python and CUDA using the Transformers framework~\cite{wolf2020transformers}. GPU HBM and DRAM were managed in blocks to improve utilization and reduce fragmentation, with dynamic allocation based on inference demands. Data transfer between GPU and DRAM was handled by a dedicated CUDA stream, and a separate thread tracked transfer progress and memory block status. Continuous batching was enabled to improve throughput.

\niparagraph{Workloads and Models.}
The experiment workload was derived from the ShareGPT dataset~\cite{ShareGPT_Vicuna_unfiltered}. Due to uneven request length distribution, we selected up to 50 samples per length group (±5\% fluctuation). Experiments used the DeepSeek-V2-Lite~\cite{DeepSeekCoderV2LiteBase, bi2024deepseek, deepseekai2024deepseekv2strongeconomicalefficient}, Qwen1.5~\cite{Qwen1.5MoEA2.7B, qwen_moe}, and Qwen2.0~\cite{Qwen2_57B_A14B_Instruct_GPTQ_Int4, qwen2} models, applying 4-bit quantization to Qwen2.0. 

To simulate resource-constrained settings, GPU memory was capped at 20GB across all platforms. The Ascend 910B lacks INT4 quantization support, preventing Qwen2.0 execution under these limits. Although runtime data for Qwen2.0 on 910B were unavailable, consistent performance trends from DeepSeek and Qwen1.5 confirm that \systemname\ effectively reduces waiting latency across devices. Once quantization is supported on 910B, the remaining experiments will be completed.

\niparagraph{Baseline. }
We compared \systemname\ with a baseline approach (Baseline) implemented using the Transformers, ProMoE~\cite{song2024promoe} and pre-gate jobs from yandex~\cite{eliseev2023fast}. We compare the results of our work with ProMoe and transformers with respect to the reduction in waiting latency as well as cache miss situation for cross-layer prediction, and the accuracy of the activated experts required for cross-layer prediction. For pre-gate aspects, such as the accuracy of pre-gate, we mainly compare our results with Yandex's work.

\subsection{Overall Latency Declines}
\label{overall_latency_declines}

\begin{figure*}[h]
    \centering
    \begin{minipage}{0.33\linewidth}
        \centering
        \subfloat[Nvidia A6000]{%
            \includegraphics[width=0.99\linewidth]{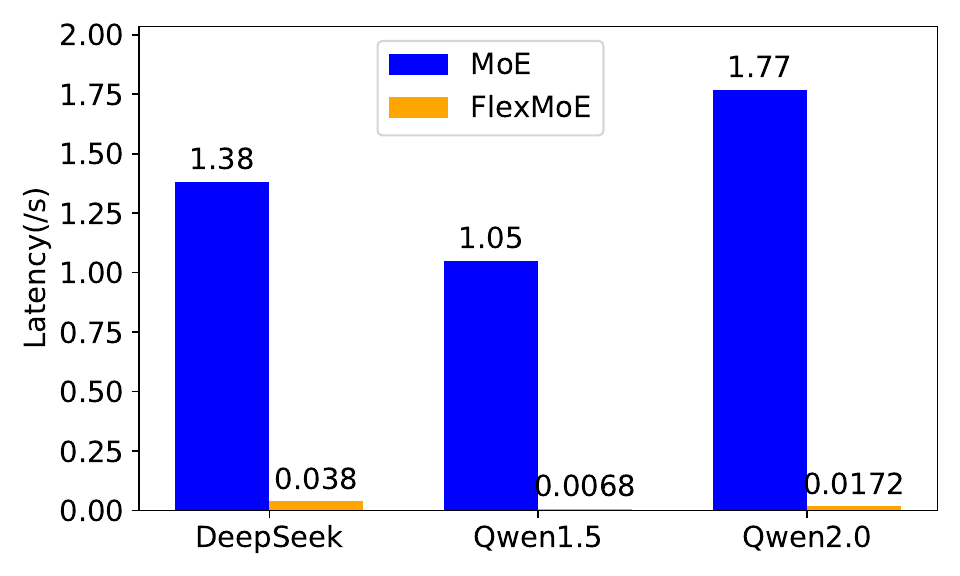}
        }
    \end{minipage}
    \begin{minipage}{0.33\linewidth}
        \centering
        \subfloat[Nvidia H20]{%
            \includegraphics[width=0.99\linewidth]{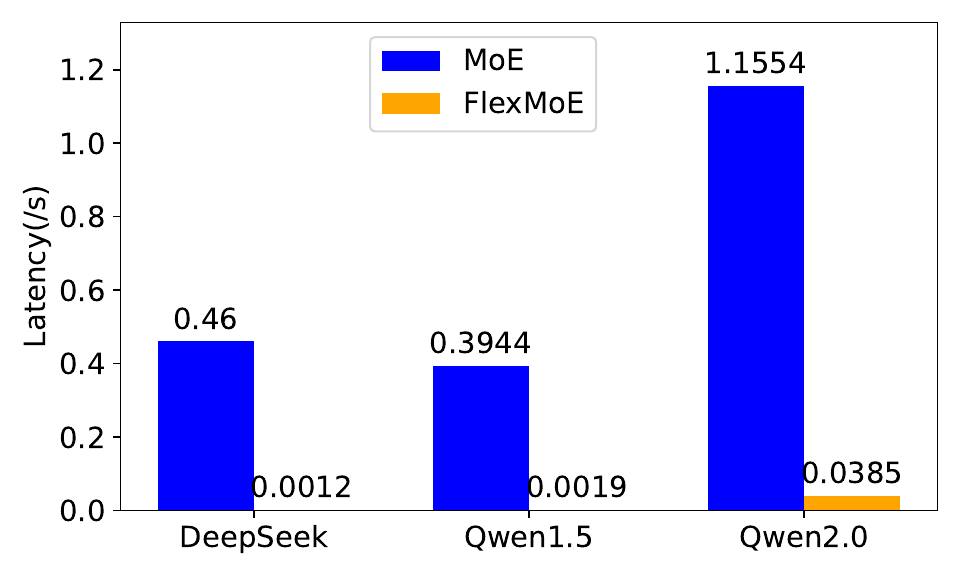}
        }
    \end{minipage}
    \begin{minipage}{0.33\linewidth}
        \centering
        \subfloat[Ascend 910B]{%
            \includegraphics[width=0.99\linewidth]{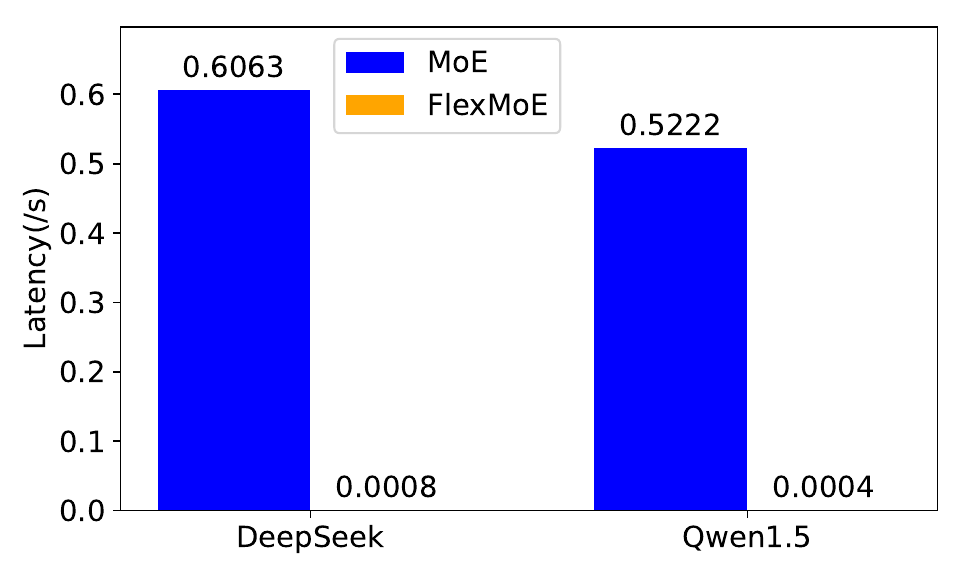}
        }
    \end{minipage}
    \vspace{-1.0em}
    \caption{Overall Waiting Latency}
    \label{fig:E2}
    \vspace{-1.0em}
\end{figure*}

In this section, we introduce the performance improvements brought by \systemname. The primary objective of \systemname\ is to dynamically predict the activation of experts with variable step sizes, enabling the prefetching of required experts and efficient management of GPU memory. This approach collaboratively reduces both waiting latency and cache miss latency. Our experiments first focus on evaluating the accuracy of expert predictions. Subsequently, as demonstrated in ~\autoref{fig:E2}, our system effectively reduces the latency encountered during the operation of MoE models.

We constructed a test dataset with prompt lengths from 8 to 1024 to evaluate three MoE models: DeepSeek, Qwen1.5, and Qwen2. As shown in the figure, relative to baseline, all models achieved a major reduction in combined waiting and cache-miss latency, averaging 98.5\%. Qwen1.5 showed the greatest gain, reducing latency by over 99.9\%.

This latency reduction highlights the effectiveness of our approach in addressing two key MoE inference bottlenecks: waiting and cache-miss latency. Waiting occurs when computation stalls for expert data, while cache misses arise when requested parameters are absent from GPU memory. Both degrade performance by causing stalls and memory-access overhead. Through dynamic prediction and memory-aware scheduling, our method proactively mitigates these effects, improving execution continuity, inference stability, and overall efficiency.

In cross-device evaluation, we observed an unexpected result: although \systemname\ reduced end-to-end latency on both H20 and A6000 platforms, total latency on the H20 was still higher, despite its superior specifications—greater compute, memory, and bandwidth—as shown in~\autoref{fig:E2} (a) and (b). The H20 provides roughly 240 TFLOPS FP16 performance versus 38.7 TFLOPS on the A6000.

We hypothesize that this occurs because H20’s memory bandwidth does not scale proportionally with its compute power. During early inference, when step size \(S\) is small, H20 finishes layer computation quickly but then idles while fetching expert parameters, increasing waiting latency. In contrast, A6000’s slower computation naturally overlaps with expert loading, reducing idle time and yielding lower observed latency under the same scheduling policy.

\subsection{Pre-Gate-Based Predictor}
\label{pre-gate-based Predictor}

In \autoref{overall_latency_declines}, we present a comparative analysis showcasing the latency reductions obtained by integrating \systemname\ into multiple MoE models relative to a standard baseline. Within the expert preloading phase of MoE inference, the precision of expert selection has a direct impact on runtime performance. Higher prediction fidelity leads to fewer delays and cache-related inefficiencies, ultimately contributing to reduced overall latency. This section focuses on evaluating the effectiveness of our predictor by analyzing its accuracy across different inference scenarios.

Since the prediction model operates solely on token-level features and expert activation logs, its accuracy is independent of the underlying GPU device. Therefore, all reported results in this section are based on evaluations conducted on the A6000 platform.

\begin{figure}
    \centering 
    \includegraphics[width=0.99\linewidth]{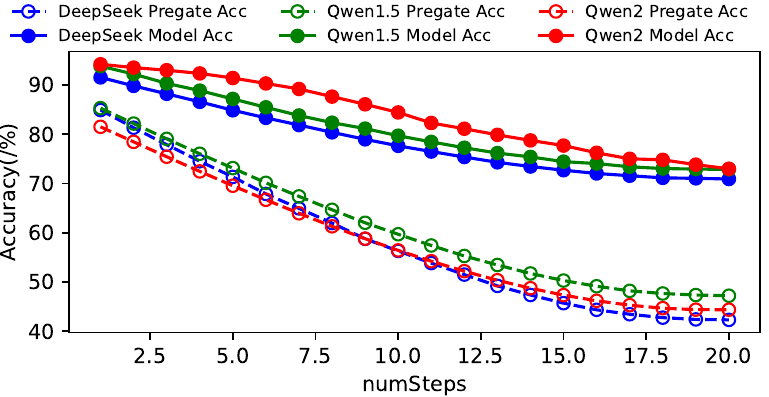}
    \captionsetup{belowskip=0em,aboveskip=0.0em}
    \vspace{-1.0em}
    \caption{Accuracy of pre-gate and Predictor}
    \label{fig:Accuracy of pre-gate and Predictor}
    \vspace{-1.0em}
\end{figure}

As shown in~\autoref{fig:Accuracy of pre-gate and Predictor}, we evaluated 410 requests of varying lengths to measure prediction accuracy when skipping layers in DeepSeek, Qwen1.5, and Qwen2.0. The trained model notably improved accuracy by an average of 21.79\% (up to 30.36\%) compared to the pre-gate method. Compared with proactive prediction~\cite{song2024promoe}, our approach achieved higher accuracy at the same step size and exhibited a smoother decay trend. 

\begin{table}[ht]
\centering
\renewcommand{\arraystretch}{1.2}
\caption{Predictor and pre-gate Accuracy Comparison}
\label{tab:accuracy_comparison}
\resizebox{0.9\linewidth}{!}{
\begin{tabular}{c c c c}
\hline\hline
\textbf{Group} & \textbf{\(c_p\)} & \textbf{\(c_g\)} & \(\Delta_\infty = c_p - c_g\) \\
\hline
DeepSeek  & 63.44 & 26.43 & 37.01 \\
Qwen1.5   & 65.31 & 33.29 & 32.02 \\
Qwen2     & 60.45 & 29.61 & 30.84 \\
\hline\hline
\end{tabular}
}
\vspace{-1.0em}
\end{table}

To model the accuracy change, both Predictor and pre-gate were fit to an exponential decay function:
\[
f(t) = a e^{-b t} + c,
\]
where \(t\) is the step number, and \(a, b, c\) denote initial deviation, decay rate, and asymptotic accuracy.  
Their fitted forms are:
\[
P(t) = a_p e^{-b_p t} + c_p, \quad G(t) = a_g e^{-b_g t} + c_g,
\]
yielding the accuracy gap:
\[
\Delta(t) = P(t) - G(t) = (a_p - a_g)e^{-b_p t} + (c_p - c_g),
\]
with the limit \(\Delta_\infty = c_p - c_g\).  
Fitted parameters \(c_p\) and \(c_g\) for each group are summarized below.

\begin{figure*}[h]
    \centering
    \begin{minipage}{0.33\linewidth}
    \centering
        \subfloat{\includegraphics[width=0.99\linewidth]{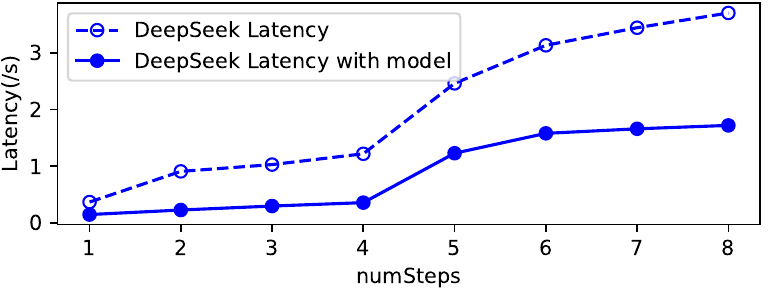}}
    \end{minipage}
    \begin{minipage}{0.33\linewidth}
    \centering
        \subfloat{\includegraphics[width=0.99\linewidth]{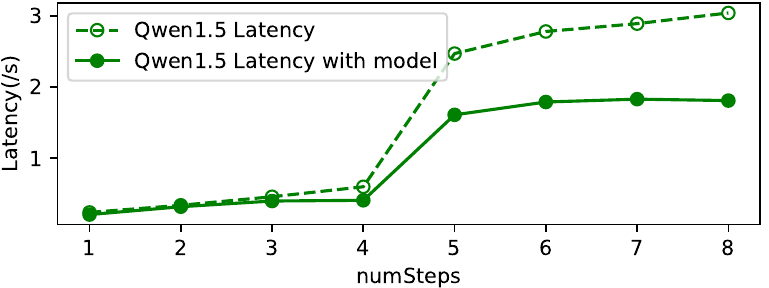}}
    \end{minipage}
    \begin{minipage}{0.33\linewidth}
    \centering
        \subfloat{\includegraphics[width=0.99\linewidth]{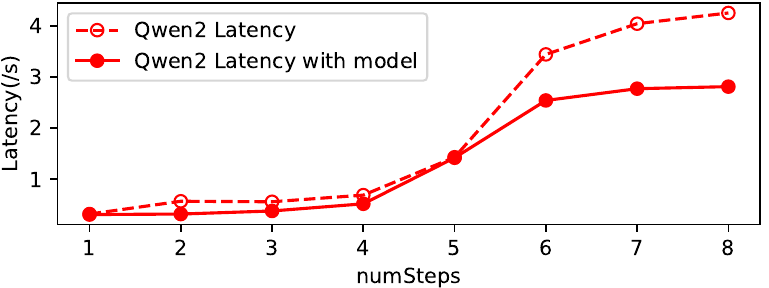}}
    \end{minipage}
    \vspace{-1.0em}
    \caption{Latency of different models with our pre-gate-based predictor}
    \vspace{-0.5em}
    \label{fig:E1_latency}
    \vspace{-0.5em}
\end{figure*}

Based on the results, the following observations can be made:
\begin{itemize}[leftmargin=1.5em, itemsep=1pt, topsep=2pt]
    \item For all mpdel groups, the difference \( \Delta_\infty > 0 \), indicating that Predictor Accuracy is consistently higher than pre-gate Accuracy.
    \item The values confirm that Predictor Accuracy has a significant and stable advantage over pre-gate Accuracy, with differences of 37.01\%, 32.0\% and 30.84\%, respectively.
\end{itemize}

Additionally, we monitored the cache miss latency in the MoE model after implementing the new prediction model. As shown in~\autoref{fig:E1_latency}, cache miss latency consistently decreased, further demonstrating that the high accuracy of our model effectively reduces caching and communication overhead in MoE.

Since the observed trends were consistent across different hardware platforms, we omit full visualizations for each device to conserve space. Instead, we present results obtained on the A6000 as a representative case. For completeness, \autoref{tab:Dynamic_predictor} summarizes the overall end-to-end latency achieved under our dynamic step size mechanism across all evaluated devices. Compared to the baseline results in~\autoref{tab:Dynamic_baseline}, it is evident that all MoE models exhibit significant latency reductions across all devices.

\begin{table}[t]
\centering
\renewcommand{\arraystretch}{1.2}
\caption{Latency of baseline on devices}
\label{tab:Dynamic_baseline}
\resizebox{0.9\linewidth}{!}{
\begin{tabular}{c   c   c   c}
\hline\hline 
\textbf{Group} & \textbf{\(A6000\)} & \textbf{\(H20\)} & \(Ascend\ 910B\) \\
\hline
DeepSeek  & 1.38 & 0.46 & 0.6063 \\
Qwen1.5   & 1.05 & 0.39 & 0.5222 \\
Qwen2     & 1.77 & 1.16 & - \\ \hline\hline
\end{tabular}
}
\vspace{-1.0em}
\end{table}

\begin{table}
\centering
\renewcommand{\arraystretch}{1.2}
\caption{Latency of models with Predictor}
\resizebox{0.9\linewidth}{!}{
\label{tab:Dynamic_predictor}
\begin{tabular}{c   c   c   c}
\hline\hline 
\textbf{Group} & \textbf{\(A6000\)} & \textbf{\(H20\)} & \(Ascend\ 910B\) \\
\hline
DeepSeek  & 0.0441 & 0.0012 & 0.0624 \\
Qwen1.5   & 0.0384 & 0.0379 & 0.0686 \\
Qwen2     & 0.1252 & 1.0176 & - \\ \hline\hline
\end{tabular}
}
\vspace{-1.0em}
\end{table}

\begin{figure*}[ht]
    \centering
    \begin{minipage}{0.33\linewidth}
    \centering
        \subfloat{\includegraphics[width=0.99\linewidth]{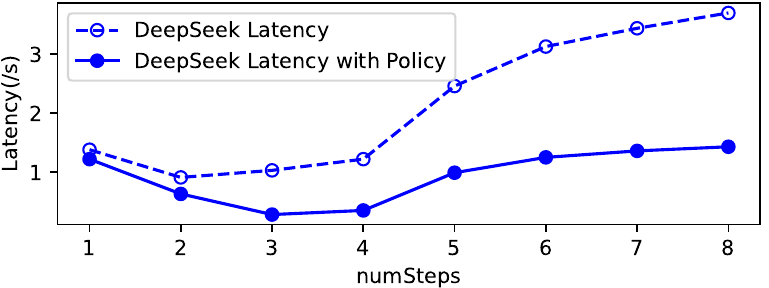}}
    \end{minipage}
    \begin{minipage}{0.33\linewidth}
    \centering
        \subfloat{\includegraphics[width=0.99\linewidth]{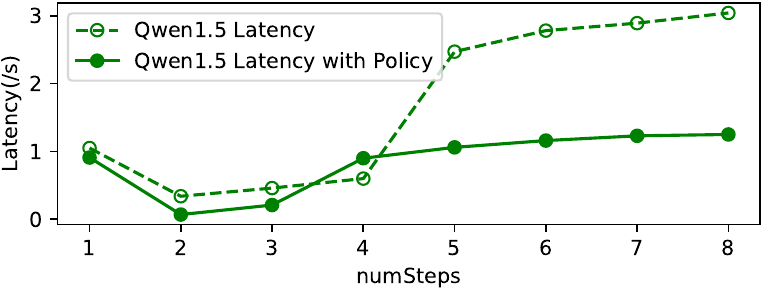}}
    \end{minipage}
    \begin{minipage}{0.33\linewidth}
    \centering
        \subfloat{\includegraphics[width=0.99\linewidth]{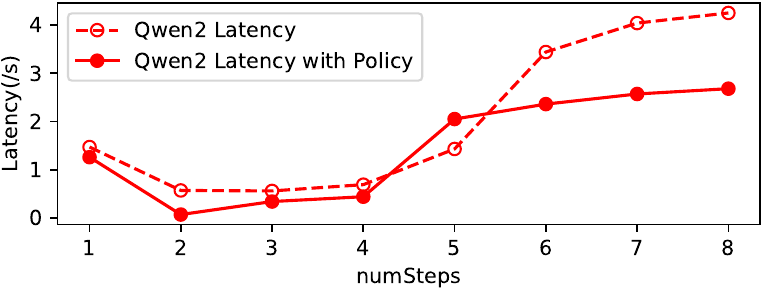}}
    \end{minipage}
    \vspace{-1.0em}
    \caption{Latency of different models with New Memory Management}
    \label{fig:E3_latency}
    \vspace{-0.5em}
\end{figure*}

\subsection{Memory Management Policy}
\label{Memory_Management_Policy}

We implemented a memory management policy based on the Least Recently Used (LRU) strategy to provide fine-grained control over expert swap operations. As shown in~\autoref{fig:E3_latency}, the MoE model with this policy consistently achieves lower latency during cross-layer prediction than the baseline without memory management.

A key observation is the sudden latency “jump” around a $S=4$, mainly caused by limited GPU memory. Smaller step sizes increase expert predictions, triggering frequent swap-in. As the model progresses, many new experts are loaded, eventually exceeding memory capacity. When this occurs, previously swapped-in experts are evicted to free space; if later required, they must be reloaded, causing sharp latency spikes. This behavior consistently appears near step size 4 on our test systems, though the exact threshold varies with GPU memory size.

\begin{figure*}[htp]
    \centering
    \begin{minipage}{0.33\linewidth}
        \centering
        \subfloat[Nvidia A6000]{%
            \includegraphics[width=0.99\linewidth]{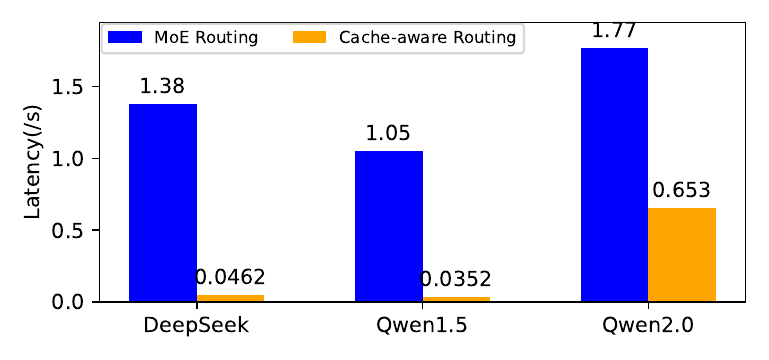}
        }
    \end{minipage}
    \begin{minipage}{0.33\linewidth}
        \centering
        \subfloat[Nvidia H20]{%
            \includegraphics[width=0.99\linewidth]{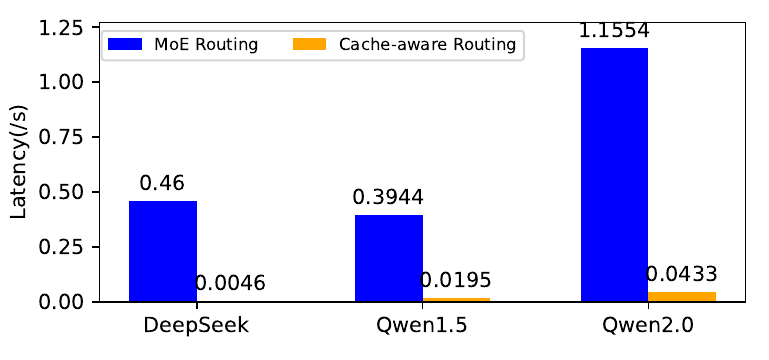}
        }
    \end{minipage}
    \begin{minipage}{0.33\linewidth}
        \centering
        \subfloat[Ascend 910B]{%
            \includegraphics[width=0.99\linewidth]{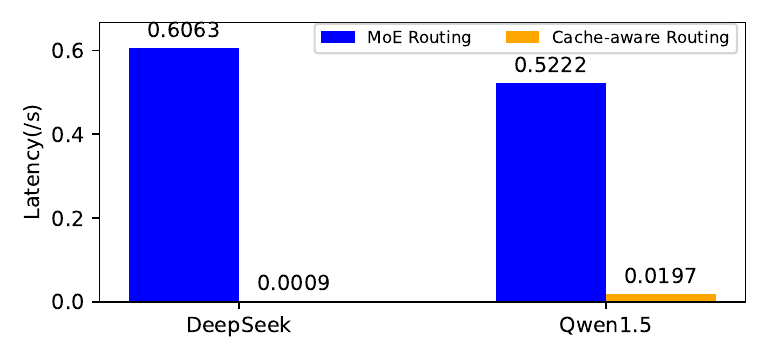}
        }
    \end{minipage}
    \vspace{-1.0em}
    \caption{Comparison of latency with or without cache-aware routing across different models}
    \label{fig:Comparison of Latency with or without Cache-ware Routing}
    \vspace{-1.0em}
\end{figure*}

As with the \autoref{pre-gate-based Predictor} evaluation, we report results based solely on the A6000 platform, since all tested devices exhibit consistent trends. In~\autoref{tab:Dynamic_CCR}, we further evaluate the effectiveness of our cache-aware routing mechanism across multiple hardware platforms. Compared to the baseline, models incorporating this mechanism show a clear reduction in end-to-end latency, demonstrating its capability to alleviate expert loading bottlenecks and reduce communication overhead in large-scale MoE inference.

\begin{table}[t]
\centering
\renewcommand{\arraystretch}{1.2}
\caption{Latency of models with Cache-aware Routing}
\resizebox{0.9\linewidth}{!}{
\label{tab:Dynamic_CCR}
\begin{tabular}{c   c   c   c}
\hline\hline 
\textbf{Group} & \textbf{\(A6000\)} & \textbf{\(H20\)} & \(Ascend\ 910B\) \\
\hline
DeepSeek  & 0.0328 & 0.0029 & 0.0020 \\
Qwen1.5   & 0.0216 & 0.0022 & 0.0018 \\
Qwen2     & 0.1034 & 0.0978 & - \\ \hline\hline
\end{tabular}
}
\vspace{-1.5em}
\end{table}

\subsection{Cache-aware Routing}
\label{Cache-aware Routing}

As shown in~\autoref{fig:Comparison of Latency with or without Cache-ware Routing}, we compared end-to-end latency in three MoE models—DeepSeek, Qwen1.5, and Qwen2.0—with and without the proposed cache-aware routing mechanism. Across all models, integrating cache-aware routing significantly reduced latency by mitigating delays from expert loading and cache misses. This improvement is essential for fast inference and efficient expert scheduling, preventing memory access overhead from becoming a bottleneck.

Quantitatively, cache-aware routing reduced latency by over 96.65\% in DeepSeek and Qwen1.5, which initially lacked cache miss mitigation and were vulnerable to latency spikes from expert swapping. Qwen2.0, designed with shared experts to ease cache inefficiencies~\cite{yang2024qwen2}, achieved a smaller yet substantial 55.58\% improvement. Even so, integrating cache-aware routing further boosted Qwen2.0’s performance, confirming that the mechanism complements existing cache optimization strategies.

\section{Related Work}
\label{sec:related_work}

\subsection{\textbf{Scalable Sparse Activation and Inference Optimization}}
Sparse activation allows MoE models to reduce computation by activating only a subset of experts per input~\cite{riquelme2021scaling,li2025uni,zhu2025exploring}. The Switch Transformer exemplifies this by routing each token to a single expert, greatly simplifying both training and inference~\cite{fedus2021switch}. Building on this principle, ExFlow exploits inter-layer expert affinity to colocate frequently co-activated experts and reduce cross-GPU communication~\cite{yao2024exploiting}, while TA-MoE incorporates network topology into its dispatch strategy for communication-aware scheduling, achieving up to 4.77$\times$ acceleration~\cite{chen2022ta}. These topology- and scheduling-aware designs effectively lower communication overhead but still face challenges under strict memory constraints. ExpertFlow addresses these limits through predictive caching and token-aware scheduling, cutting GPU memory usage by 93.72\% and boosting throughput by 9.99$\times$~\cite{he2024expertflow}. Similarly, MoE-Lightning employs hierarchical memory management for dynamic load balancing, achieving 10.3$\times$ speedup~\cite{cao2024moe}. MoE-ERAS improves efficiency by prioritizing experts already resident in fast memory~\cite{bambhaniya2024moe}, AdaMOE adapts expert activation per token via null experts~\cite{zeng2024adamoe}, and HOBBIT leverages mixed-precision offloading with quantized (e.g., Int4) experts, layer-wise prefetching, and multi-level caching~\cite{tang2024hobbit}. Collectively, these frameworks enhance MoE inference through memory-aware scheduling, adaptive routing, and expert approximation. \systemname\ focuses on adaptive cross-layer prefetching through a dynamic step-size predictor guided by token ID, pre-gate outputs, and expert history, minimizing cache misses and waiting latency without loss of precision.

\subsection{\textbf{Routing, Load Balancing, \& Application Adaptability}}
Effective routing is critical to MoE performance~\cite{huang2025dynamic, zhou2025moe}. Expert Choice enables load balancing by allowing experts to self-select tokens without auxiliary loss~\cite{zhou2022mixture}. AdaMOE adjusts active expert counts based on token complexity~\cite{zeng2024adamoe}, while ExFlow reduces communication by maintaining consistent token-to-expert mappings across layers~\cite{he2024expertflow}. MoDSE improves compute efficiency by assigning smaller experts to simple tokens and larger ones to complex tokens~\cite{sun2024mixture}. Together, these approaches explore dynamic expert allocation based on token characteristics to balance load and maximize parallelism. By adapting routing to token-level diversity, they reduce bottlenecks and improve throughput. Such mechanisms also help MoE systems adapt to varying workloads, enhancing inference efficiency under dynamic input distributions. Combined with memory-aware execution, these optimizations form the basis for adaptive and efficient MoE pipelines.

Beyond general-purpose language modeling, MoE architectures extend to specialized domains. TIMEMOE uses sparse activation for multitask time-series forecasting, improving accuracy across benchmarks~\cite{shi2024time}. In speech deepfake detection, dynamic gating enhances generalization to multilingual and unseen data~\cite{negroni2024leveraging}. KAMoE applies Kolmogorov-Arnold-based gating for interpretable, efficient expert selection~\cite{inzirillo2024gated}, while MOE-NP advances graph learning by assigning experts based on local and global node patterns~\cite{shi2024mixture}. These domain-specific adaptations show how routing and gating can be customized for task-dependent data, expanding MoE’s versatility in real-world applications. Overall, they emphasize the need for routing frameworks that are both computationally efficient and adaptable to diverse learning scenarios.
\section{Conclusion}
\label{sec:conclusion}
The proposed system, \systemname, is a dynamic prefetching framework for adaptive cross-layer prediction in MoE models, integrating flexible memory management and cache-aware routing to reduce expert loading and waiting latency. It combines a dynamic layer-adjustment algorithm, expert-priority-based memory management, and real-time routing under a first-come, first-served protocol. Experiments show that \systemname\ improves expert prediction accuracy by over 30\% and eliminates up to 99.9\% of waiting latency while maintaining stable performance across different hardware and workloads, demonstrating strong scalability for real-world MoE inference.
\nocite{langley00}

\bibliographystyle{mlsys2025}
\bibliography{references}

\begin{thebibliography}{42}
\providecommand{\natexlab}[1]{#1}
\providecommand{\url}[1]{\texttt{#1}}
\expandafter\ifx\csname urlstyle\endcsname\relax
  \providecommand{\doi}[1]{doi: #1}\else
  \providecommand{\doi}{doi: \begingroup \urlstyle{rm}\Url}\fi

\bibitem[AI()]{DeepSeekCoderV2LiteBase}
AI, D.
\newblock Deepseek-coder-v2-lite-base.
\newblock \url{https://huggingface.co/deepseek-ai/DeepSeek-Coder-V2-Lite-Base}.

\bibitem[Anonymous()]{ShareGPT_Vicuna_unfiltered}
Anonymous.
\newblock Sharegpt vicuna unfiltered.
\newblock \url{https://huggingface.co/datasets/anon8231489123/ShareGPT_Vicuna_unfiltered}.

\bibitem[Bambhaniya et~al.(2024)Bambhaniya, Kumar, and Krishna]{bambhaniya2024moe}
Bambhaniya, A.~R., Kumar, S.~C., and Krishna, T.
\newblock Moe-eras: Expert residency aware selection.
\newblock In \emph{Machine Learning for Computer Architecture and Systems 2024}, 2024.

\bibitem[Bi et~al.(2024)Bi, Chen, Chen, Chen, Dai, Deng, Ding, Dong, Du, Fu, et~al.]{bi2024deepseek}
Bi, X., Chen, D., Chen, G., Chen, S., Dai, D., Deng, C., Ding, H., Dong, K., Du, Q., Fu, Z., et~al.
\newblock Deepseek llm: Scaling open-source language models with longtermism.
\newblock \emph{arXiv preprint arXiv:2401.02954}, 2024.

\bibitem[Brown et~al.(2020)Brown, Mann, Ryder, Subbiah, Kaplan, Dhariwal, Neelakantan, Shyam, Sastry, Askell, et~al.]{brown2020language}
Brown, T., Mann, B., Ryder, N., Subbiah, M., Kaplan, J.~D., Dhariwal, P., Neelakantan, A., Shyam, P., Sastry, G., Askell, A., et~al.
\newblock Language models are few-shot learners.
\newblock \emph{Advances in neural information processing systems}, 33:\penalty0 1877--1901, 2020.

\bibitem[Cao et~al.(2025)Cao, Liu, Griggs, Schafhalter, Liu, Sheng, Gonzalez, Zaharia, and Stoica]{cao2025moe}
Cao, S., Liu, S., Griggs, T., Schafhalter, P., Liu, X., Sheng, Y., Gonzalez, J.~E., Zaharia, M., and Stoica, I.
\newblock Moe-lightning: High-throughput moe inference on memory-constrained gpus.
\newblock In \emph{Proceedings of the 30th ACM International Conference on Architectural Support for Programming Languages and Operating Systems, Volume 1}, pp.\  715--730, 2025.

\bibitem[Cao et~al.(2024)]{cao2024moe}
Cao, S. et~al.
\newblock Moe-lightning: High-throughput moe inference on memory-constrained gpus.
\newblock \emph{arXiv preprint arXiv:2411.11217}, 2024.

\bibitem[Chen et~al.(2022)Chen, Li, Wu, Yu, and Yang]{chen2022ta}
Chen, C., Li, M., Wu, Z., Yu, D., and Yang, C.
\newblock Ta-moe: Topology-aware large scale mixture-of-expert training.
\newblock \emph{Advances in Neural Information Processing Systems}, 35:\penalty0 22173--22186, 2022.

\bibitem[DeepSeek-AI et~al.(2024)]{deepseekai2024deepseekv2strongeconomicalefficient}
DeepSeek-AI et~al.
\newblock Deepseek-v2: A strong, economical, and efficient mixture-of-experts language model, 2024.
\newblock URL \url{https://arxiv.org/abs/2405.04434}.

\bibitem[Eliseev \& Mazur(2023)Eliseev and Mazur]{eliseev2023fast}
Eliseev, A. and Mazur, D.
\newblock Fast inference of mixture-of-experts language models with offloading.
\newblock \emph{arXiv preprint arXiv:2312.17238}, 2023.

\bibitem[Fedus et~al.(2021)Fedus, Zoph, and Shazeer]{fedus2021switch}
Fedus, W., Zoph, B., and Shazeer, N.
\newblock Switch transformers: Scaling to trillion parameter models with simple and efficient sparsity, 2021.

\bibitem[Fedus et~al.(2022)Fedus, Zoph, and Shazeer]{fedus2022switch}
Fedus, W., Zoph, B., and Shazeer, N.
\newblock Switch transformers: Scaling to trillion parameter models with simple and efficient sparsity.
\newblock \emph{Journal of Machine Learning Research}, 23\penalty0 (120):\penalty0 1--39, 2022.

\bibitem[Graves(2016)]{graves2016adaptive}
Graves, A.
\newblock Adaptive computation time for recurrent neural networks.
\newblock \emph{arXiv preprint arXiv:1603.08983}, 2016.

\bibitem[He et~al.(2024)He, Zhang, Wang, Yin, Zeng, Shi, Tang, Chu, Tsang, and Soon]{he2024expertflow}
He, X., Zhang, S., Wang, Y., Yin, H., Zeng, Z., Shi, S., Tang, Z., Chu, X., Tsang, I., and Soon, O.~Y.
\newblock Expertflow: Optimized expert activation and token allocation for efficient mixture-of-experts inference.
\newblock \emph{arXiv preprint arXiv:2410.17954}, 2024.

\bibitem[Huang et~al.(2025)Huang, Lu, Shan, Qu, Zhang, Guan, Hong, and Li]{huang2025dynamic}
Huang, H., Lu, S., Shan, Y., Qu, H., Zhang, F., Guan, W., Hong, Q., and Li, L.
\newblock Dynamic language group-based moe: Enhancing code-switching speech recognition with hierarchical routing.
\newblock In \emph{ICASSP 2025-2025 IEEE International Conference on Acoustics, Speech and Signal Processing (ICASSP)}, pp.\  1--5. IEEE, 2025.

\bibitem[Inzirillo \& Genet(2024)Inzirillo and Genet]{inzirillo2024gated}
Inzirillo, H. and Genet, R.
\newblock A gated residual kolmogorov-arnold networks for mixtures of experts.
\newblock \emph{arXiv preprint arXiv:2409.15161}, 2024.

\bibitem[Jordan \& Jacobs(1993)Jordan and Jacobs]{716791}
Jordan, M. and Jacobs, R.
\newblock Hierarchical mixtures of experts and the em algorithm.
\newblock In \emph{Proceedings of 1993 International Conference on Neural Networks (IJCNN-93-Nagoya, Japan)}, volume~2, pp.\  1339--1344 vol.2, 1993.
\newblock \doi{10.1109/IJCNN.1993.716791}.

\bibitem[Lepikhin et~al.(2020)Lepikhin, Lee, Xu, Chen, Vinyals, Huang, and Zoph]{lepikhin2020gshard}
Lepikhin, D., Lee, Y., Xu, Y., Chen, M., Vinyals, O., Huang, Y., and Zoph, B.
\newblock Gshard: Scaling giant models with conditional computation and automatic sharding, 2020.

\bibitem[Li et~al.(2025)Li, Jiang, Hu, Wang, Zhong, Luo, Ma, and Zhang]{li2025uni}
Li, Y., Jiang, S., Hu, B., Wang, L., Zhong, W., Luo, W., Ma, L., and Zhang, M.
\newblock Uni-moe: Scaling unified multimodal llms with mixture of experts.
\newblock \emph{IEEE Transactions on Pattern Analysis and Machine Intelligence}, 2025.

\bibitem[Negroni et~al.(2024)Negroni, Salvi, Mezza, Bestagini, and Tubaro]{negroni2024leveraging}
Negroni, V., Salvi, D., Mezza, A.~I., Bestagini, P., and Tubaro, S.
\newblock Leveraging mixture of experts for improved speech deepfake detection.
\newblock \emph{arXiv preprint arXiv:2409.16077}, 2024.

\bibitem[{NVIDIA Corporation}(2024)]{nvidiaA6000datasheet}
{NVIDIA Corporation}.
\newblock {NVIDIA RTX A6000 Datasheet}.
\newblock \url{https://resources.nvidia.com/en-us-briefcase-for-datasheets/proviz-print-nvidia-1?ncid=no-ncid}, 2024.
\newblock Accessed: 2025-06-14.

\bibitem[Pedregosa et~al.(2011)Pedregosa, Varoquaux, Gramfort, Michel, Thirion, Grisel, Blondel, Prettenhofer, Weiss, Dubourg, Vanderplas, Passos, Cournapeau, Brucher, Perrot, and Duchesnay]{scikit-learn}
Pedregosa, F., Varoquaux, G., Gramfort, A., Michel, V., Thirion, B., Grisel, O., Blondel, M., Prettenhofer, P., Weiss, R., Dubourg, V., Vanderplas, J., Passos, A., Cournapeau, D., Brucher, M., Perrot, M., and Duchesnay, E.
\newblock Scikit-learn: Machine learning in {P}ython.
\newblock \emph{Journal of Machine Learning Research}, 12:\penalty0 2825--2830, 2011.

\bibitem[Qwen({\natexlab{a}})]{Qwen1.5MoEA2.7B}
Qwen.
\newblock Qwen1.5-moe-a2.7b.
\newblock \url{https://huggingface.co/Qwen/Qwen1.5-MoE-A2.7B}, {\natexlab{a}}.

\bibitem[Qwen({\natexlab{b}})]{Qwen2_57B_A14B_Instruct_GPTQ_Int4}
Qwen.
\newblock Qwen2-57b-a14b-instruct-gptq-int4.
\newblock \url{https://huggingface.co/Qwen/Qwen2-57B-A14B-Instruct-GPTQ-Int4}, {\natexlab{b}}.

\bibitem[Riquelme et~al.(2021)]{riquelme2021scaling}
Riquelme, C. et~al.
\newblock Scaling vision with sparse mixture of experts.
\newblock \emph{Advances in Neural Information Processing Systems}, 34:\penalty0 8583--8595, 2021.

\bibitem[Shazeer et~al.(2017)Shazeer, Mirhoseini, Maziarz, Davis, Le, Hinton, and Dean]{shazeer2017outrageously}
Shazeer, N., Mirhoseini, A., Maziarz, K., Davis, A., Le, Q.~V., Hinton, G., and Dean, J.
\newblock Outrageously large neural networks: The sparsely-gated mixture-of-experts layer, 2017.

\bibitem[Shi et~al.(2024{\natexlab{a}})Shi, Wang, Nie, Li, Ye, Wen, and Jin]{shi2024time}
Shi, X., Wang, S., Nie, Y., Li, D., Ye, Z., Wen, Q., and Jin, M.
\newblock Time-moe: Billion-scale time series foundation models with mixture of experts.
\newblock \emph{arXiv preprint arXiv:2409.16040}, 2024{\natexlab{a}}.

\bibitem[Shi et~al.(2024{\natexlab{b}})Shi, Wang, Lang, Zhang, Dong, and Li]{shi2024mixture}
Shi, Y., Wang, Y., Lang, W., Zhang, J., Dong, P., and Li, A.
\newblock Mixture of experts for node classification.
\newblock \emph{arXiv preprint arXiv:2412.00418}, 2024{\natexlab{b}}.

\bibitem[Song et~al.(2024)Song, Zhong, and Chen]{song2024promoe}
Song, X., Zhong, Z., and Chen, R.
\newblock Promoe: Fast moe-based llm serving using proactive caching.
\newblock \emph{arXiv preprint arXiv:2410.22134}, 2024.

\bibitem[Sun et~al.(2024)Sun, Liu, Luan, Gao, and Wang]{sun2024mixture}
Sun, M., Liu, W., Luan, J., Gao, P., and Wang, B.
\newblock Mixture of diverse size experts.
\newblock \emph{arXiv preprint arXiv:2409.12210}, 2024.

\bibitem[Tang et~al.(2024)Tang, Liu, Hou, Pu, Wang, Heng, Li, and Guo]{tang2024hobbit}
Tang, P., Liu, J., Hou, X., Pu, Y., Wang, J., Heng, P.-A., Li, C., and Guo, M.
\newblock Hobbit: A mixed precision expert offloading system for fast moe inference.
\newblock \emph{arXiv preprint arXiv:2411.01433}, 2024.

\bibitem[Team(2024)]{qwen_moe}
Team, Q.
\newblock Qwen1.5-moe: Matching 7b model performance with 1/3 activated parameters", February 2024.
\newblock URL \url{https://qwenlm.github.io/blog/qwen-moe/}.

\bibitem[Team et~al.(2024)]{qwen2}
Team, Q. et~al.
\newblock Qwen2 technical report.
\newblock \emph{arXiv preprint arXiv:2407.10671}, 2\penalty0 (3), 2024.

\bibitem[Wang et~al.(2018)Wang, Yu, Dou, Darrell, and Gonzalez]{wang2018skipnet}
Wang, X., Yu, F., Dou, Z.-Y., Darrell, T., and Gonzalez, J.~E.
\newblock Skipnet: Learning dynamic routing in convolutional networks.
\newblock In \emph{Proceedings of the European conference on computer vision (ECCV)}, pp.\  409--424, 2018.

\bibitem[Wolf et~al.(2020)Wolf, Debut, Sanh, Chaumond, Delangue, Moi, Cistac, Rault, Louf, Funtowicz, et~al.]{wolf2020transformers}
Wolf, T., Debut, L., Sanh, V., Chaumond, J., Delangue, C., Moi, A., Cistac, P., Rault, T., Louf, R., Funtowicz, M., et~al.
\newblock Transformers: State-of-the-art natural language processing.
\newblock In \emph{Proceedings of the 2020 conference on empirical methods in natural language processing: system demonstrations}, pp.\  38--45, 2020.

\bibitem[Yang et~al.(2024{\natexlab{a}})Yang, Yang, Zhang, Hui, Zheng, Yu, Li, Liu, Huang, Wei, et~al.]{yang2024qwen2}
Yang, A., Yang, B., Zhang, B., Hui, B., Zheng, B., Yu, B., Li, C., Liu, D., Huang, F., Wei, H., et~al.
\newblock Qwen2. 5 technical report.
\newblock \emph{arXiv preprint arXiv:2412.15115}, 2024{\natexlab{a}}.

\bibitem[Yang et~al.(2024{\natexlab{b}})Yang, Qi, Gu, Wang, Gao, and Xu]{yang2024xmoe}
Yang, Y., Qi, S., Gu, W., Wang, C., Gao, C., and Xu, Z.
\newblock Xmoe: Sparse models with fine-grained and adaptive expert selection.
\newblock In \emph{Findings of the Association for Computational Linguistics ACL 2024}, pp.\  11664--11674, 2024{\natexlab{b}}.

\bibitem[Yao et~al.(2024)Yao, Anthony, Shafi, Subramoni, and Panda]{yao2024exploiting}
Yao, J., Anthony, Q., Shafi, A., Subramoni, H., and Panda, D. K.~D.
\newblock Exploiting inter-layer expert affinity for accelerating mixture-of-experts model inference.
\newblock In \emph{2024 IEEE International Parallel and Distributed Processing Symposium (IPDPS)}, pp.\  915--925. IEEE, 2024.

\bibitem[Zeng et~al.(2024)Zeng, Miao, Gao, Zhang, and Deng]{zeng2024adamoe}
Zeng, Z., Miao, Y., Gao, H., Zhang, H., and Deng, Z.
\newblock Adamoe: Token-adaptive routing with null experts for mixture-of-experts language models.
\newblock \emph{arXiv preprint arXiv:2406.13233}, 2024.

\bibitem[Zhou et~al.(2025)Zhou, Wang, Huang, Huang, Han, Feng, Deng, Luo, and Chen]{zhou2025moe}
Zhou, H., Wang, Z., Huang, S., Huang, X., Han, X., Feng, J., Deng, C., Luo, W., and Chen, J.
\newblock Moe-lpr: Multilingual extension of large language models through mixture-of-experts with language priors routing.
\newblock In \emph{Proceedings of the AAAI Conference on Artificial Intelligence}, volume~39, pp.\  26092--26100, 2025.

\bibitem[Zhou et~al.(2022)Zhou, Lei, Liu, Du, Huang, Zhao, Dai, Le, Laudon, et~al.]{zhou2022mixture}
Zhou, Y., Lei, T., Liu, H., Du, N., Huang, Y., Zhao, V., Dai, A.~M., Le, Q.~V., Laudon, J., et~al.
\newblock Mixture-of-experts with expert choice routing.
\newblock \emph{Advances in Neural Information Processing Systems}, 35:\penalty0 7103--7114, 2022.

\bibitem[Zhu et~al.(2025)Zhu, Yang, Zheng, Xu, Shi, Zhang, Chen, and Shen]{zhu2025exploring}
Zhu, J., Yang, C., Zheng, K., Xu, Y., Shi, Z., Zhang, Y., Chen, Q., and Shen, Y.
\newblock Exploring sparse moe in gans for text-conditioned image synthesis.
\newblock In \emph{Proceedings of the Computer Vision and Pattern Recognition Conference}, pp.\  18411--18423, 2025.

\end{thebibliography}

\end{document}